# Reciprocity-induced symmetry in the round-trip transmission through complex systems

SZU-YU LEE,[1,2] VICENTE J. PAROT,[1] BRETT E. BOUMA,[1,2,3] AND MARTIN VILLIGER[1,*]

[1]Harvard Medical School and Massachusetts General Hospital, Wellman Center for Photomedicine, Boston, MA 02114, USA
[2] Harvard-MIT Health Sciences and Technology, Massachusetts Institute of Technology, Cambridge, MA 02139, USA
[3] Institute for Medical Engineering and Science, Massachusetts Institute of Technology, Cambridge, MA 02139, USA

*Corresponding author: *mvilliger@mgh.harvard.edu

**Abstract:**

Reciprocity is a fundamental principle of wave physics and directly relates to the symmetry in the transmission through a system when interchanging the input and output. The coherent transmission matrix (TM) is a convenient method to characterize wave transmission through general media. Here we demonstrate the optical reciprocal nature of complex media by exploring their TM properties. We measured phase-corrected TMs of forward and round-trip propagation through a looped 1m-long step-index optical multimode fiber (MMF) to experimentally verify a transpose relationship between forward and backward transmission. This symmetry impedes straightforward MMF calibration from proximal measurements of the round-trip TM. Furthermore, we show how focusing through the MMF with digital optical phase conjugation is compromised by system loss, since time reversibility relies on power conservation. These insights may inform development of new imaging techniques through complex media and coherent control of waves in photonic systems.

## I. Introduction

The bi-directional transmission through photonic systems is governed by the universal Lorentz reciprocity (or Helmholtz reciprocity), which states that light propagating along a reversed path experiences the exact same transmission coefficient as in the forward direction, independent of the path complexity [1,2] or the presence of loss [3–5]. In the linear regime, this suggests a definite relation, or symmetry, between the forward and the backward transmission when interchanging the source and detector. This symmetry not only underlies the behavior of common optical components like polarizers, beam-splitters, and wave-plates, but also engenders surprising physical phenomena in complex systems such as coherent backscattering (or weak localization) and Anderson localization [6,7]. Optical phase conjugation is a well-known consequence of this symmetry in loss-free systems, whereby an original light distribution is replicated by reversing the propagation direction of the detected field while conjugating its wave-front. Digital optical phase conjugation (DOPC) has been well established for focusing and imaging through complex or disordered media, including multimode fibers (MMFs) [8–11]. However, the more general underlying transmission symmetry of bi-directional light transmission through complex systems and its implications have not been explicitly demonstrated and discussed.

Light transmission through MMFs is typically considered a chaotic process, as the modal scrambling results in the generation of random speckle patterns at the output [12]. MMFs are of particular interest for studying optical transport in complex media due to their finite degrees of freedom (DOF) and optical energy confinement. Owing to their high data transmission capacity in an ultra-small footprint, MMFs have gained significant attention and hold great promise for optical communication and biomedical endoscopic applications [13,14]. For instance, once measured, the apparently chaotic transmission through a MMF can be harnessed to relay image information from the distal to the proximal fiber end, enabling the reconstruction of an image of the distal object [15–17]. Nevertheless, imaging through MMFs remains technically challenging, and exploring fundamental properties of MMF light transmission may help develop new strategies to advance MMF-based imaging and endoscopy [18].

In the present study, we investigate MMF transmission properties using a monochromatic coherent transmission matrix (TM) formalism [19,20] and experimentally demonstrate the transpose symmetry between the forward and backward TMs in this complex medium imposed by general optical reciprocity. The TM description is a subpart of the common scattering matrix formalism [1,21,22] and offers a simpler framework that decouples the input and output channels. We also show that while DOPC enables focusing through MMF, the focusing performance declines with increasing loss since the time reversibility is corrupted when power is not conserved. Finally, we discuss the implications of the resulting transpose symmetry on calibrating MMF transmission with only access to the proximal

## II. Results

### 2.1 Measuring the single-pass forward $T_{fw}$ and double-pass round-trip $T_{2X}$ of a MMF

As shown in Fig. 1, the optical transmission through a general medium from an input surface to an output surface can be expressed by a TM, where each element is a complex coefficient specifying the amplitude and phase evolution of the transmitted monochromatic field between the corresponding pair of input and output spatial channels. The spatial channels correspond to the sampling locations on the input and output surfaces, respectively, and are assumed to be sufficiently dense to correctly sample the electromagnetic fields. We can then express forward light transmission from the proximal to distal end as

$$\vec{t} = T_{fw} \times \vec{s}, \tag{1}$$

where $\vec{t}$ and $\vec{s}$ are vectorized representations of the distal output field and proximal input field, respectively. The backward light transmission from the distal to proximal end can be written as

$$\vec{u} = T_{bw} \times \vec{t}, \tag{2}$$

where $\vec{u}$ and $\vec{t}$ are the proximal output field and distal input field, respectively. According to the reciprocity theorem, light propagating along the reversed path between input and output will experience the same transmission coefficient as in the forward direction. In the context of Jones matrices, which describe the relation between the polarization states of the input and the output field propagating through an optical system, de Hoop's notion of reciprocity manifests as a transpose relationship between the Jones matrices describing forward and reverse transmissions [2]. By analogy with the Jones matrix formalism, when interchanging the input and output spatial channels of the medium, reciprocity instructs that $T_{bw}$ is the transpose of $T_{fw}$:

$$T_{bw} = T_{fw}^T, \tag{3}$$

where the superscript T indicates the regular matrix transpose. Also, since sequential light transmission is modeled as TM multiplication, the round-trip transmission through the same medium, $T_{2X}$ (light transmits to and is reflected from the distal side, then travels back to the proximal side) equals the product of $T_{bw}$ and $T_{fw}$:

$$T_{2X} = T_{bw} \times T_{fw} = T_{fw}^T \times T_{fw}, \tag{4}$$

making $T_{2X}$ a transpose symmetric matrix,

$$T_{2X} = T_{2X}^T. \tag{5}$$

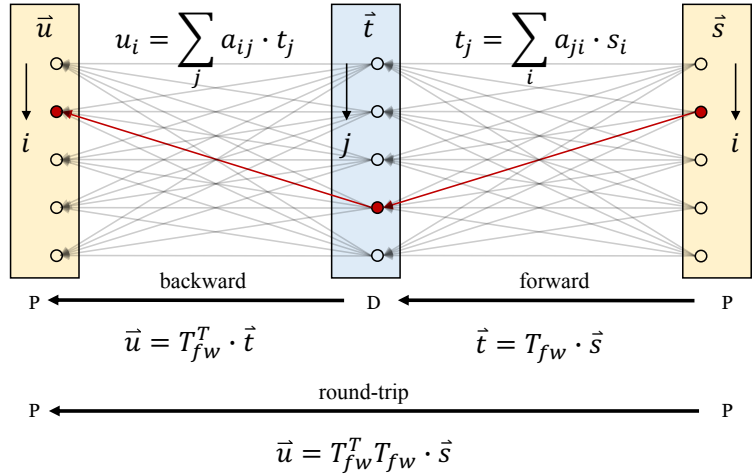

**Fig. 1.** Schematic of forward and backward TMs characterizing transmission between the proximal (P) and distal (D) ends of a linear optical system. The round-trip transmission from and to the proximal end is unfolded to reveal the hidden transpose symmetry when flipping the direction of an optical path (gray arrows) linking a pair of spatial channels. The vectors $\vec{u}$, $\vec{s}$, and $\vec{t}$ represent complex fields with constituent spatial channels indexed by $i$ and $j$ on the proximal and distal ends, respectively. Each element $a_{ji}$ of the forward TM describes the complex contribution of proximal input channel $i$ to distal output channel $j$. The red arrows link a pair of spatial channels in the forward and backward transmission. Owing to reciprocity, both directions feature the same transmission coefficient, yet correspond to transposed elements in the corresponding TMs, with interchanged row and column indices.

To experimentally verify Eqs. 3 and 4, we measured the monochromatic TMs, $T_{fw}$ and $T_{2X}$, of a 1-m-long MMF randomly coiled with a minimum radius of curvature of 23 mm, using the setup shown in Fig. 2. A laser beam ($\lambda$ = 1550 nm, linewidth < 100 kHz) was reflected on a phase-only spatial light modulator (SLM, Model P1920-850-1650-HDMI, Meadowlark Optics) in a vertical (V) polarization state, and then focused by an objective lens (Mitutoyo Plan Apo NIR Infinity Corrected) with a numerical aperture (NA) of 0.4 into a 2.5 μm full-width at half maximum (FWHM) spot on the facet of the step-index MMF with 105 μm core diameter and a NA of 0.22 (FG105LCA, Thorlabs). The MMF theoretically supports ∼550 guided modes per linear polarization [23]. The angular spectrum of the spot exceeded the NA of the MMF to ensure efficient population of high-order modes. To measure $T_{fw}$, the forward TM (Fig. 2(a)), we coupled the focal spot into the MMF through input spatial channels on the proximal side, and imaged the speckle pattern exiting from output spatial channels on the distal side with another identical objective and a tube lens (f = 30 cm) onto an InGaAs camera (OW1.7-VS-CL-LP-640, Raptor Photonics) with a vertically oriented, linear polarizer (LP) placed in front of it. A tilted plane reference wave, polarized by the same polarizer, interfered with the speckle pattern to record the complex image of the speckle pattern field through off-axis holography in the V polarization state. To release digital storage burden, we down-sampled the complex image at a defined grid of 2637 positions. To uniformly probe all MMF guided modes, this procedure was repeated in an oversampling fashion for a dense grid of 695 equally spaced illuminating foci sequentially generated by phase gradients on the SLM. Rearranging column by column the ensemble of vectorized complex output images recorded over all input spatial channels constructed the $T_{fw}$ representing the linear transformation of light traveling from the proximal facet to the distal facet. Due to the difference in the number of input and output sampling positions, the $T_{fw}$ is a tall rectangular matrix.

For measuring $T_{2X}$, the double-pass TM, as shown in Fig. 2(b), we again sequentially coupled light into the MMF from the proximal end through the same set of input spatial channels. On the distal side, we replaced the camera used for measuring in the forward transmission with a gold-coated mirror to reflect the light back into the MMF. The same V linear polarizer, previously in front of the camera and now in front of the gold-coated mirror, was necessary to avoid polarization cross-talk and achieve identical forward light propagation paths in both single-pass and double-pass scenarios. This is because the spatial and polarization DOF are coupled through mode mixing during light propagation in the MMF, and the MMF output polarization states are generally different from the input polarization state [24]. On the proximal side, we recorded the round-trip transmission by decoupling its path from the illumination with a non-polarizing beam-splitter. To preserve the symmetry between the illumination and the detection configurations and to obtain a square matrix $T_{2X}$, we sampled the recorded output fields at the 695 positions defined by the input focus positions. Furthermore, to mitigate specular reflections at both the distal and proximal facets, wedge prism mounting shims (SM1W1122, Thorlabs) filled with index-matching gel (G608N3, Thorlabs) were used to cover both facets for measurements of forward and double-pass TMs.

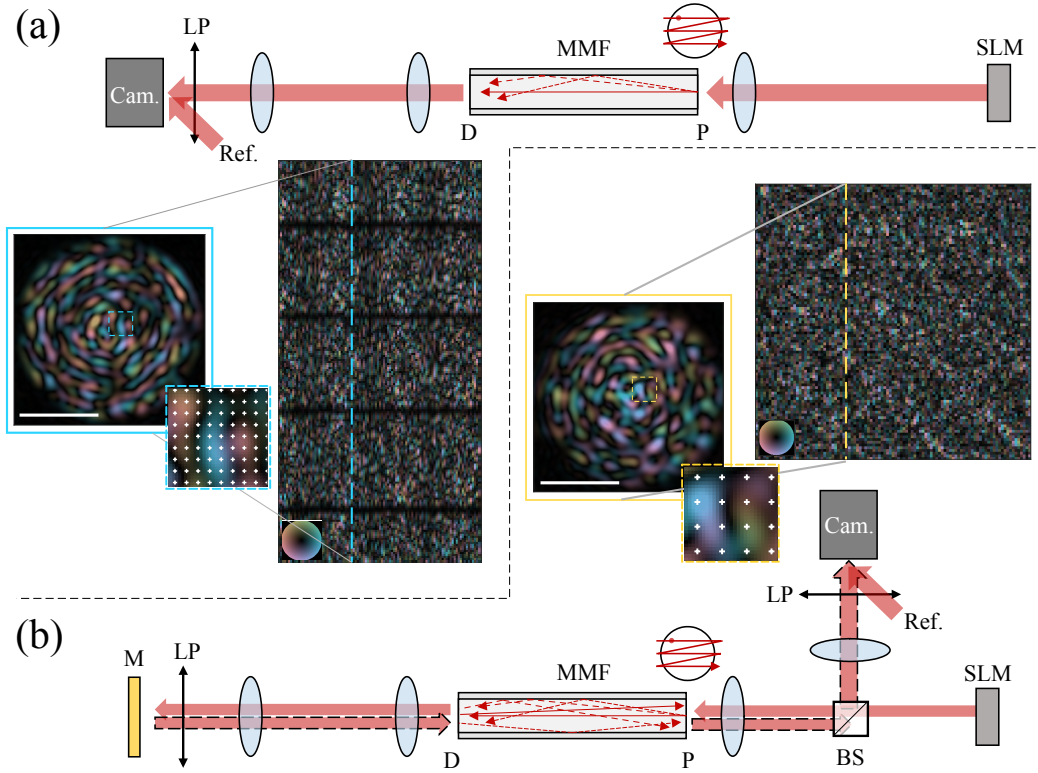

**Fig. 2.** Measurements of the MMF TMs. The fiber, though drawn as if it were straight, was in fact coiled in experiments. P: proximal. D: distal. LP: linear polarizer. Ref.: reference wave. Cam.: camera. M: gold-coated mirror. BS: beam splitter. (a) A focus was scanned by the SLM across 695 positions distributed over the MMF proximal facet. The output light field interfered with a reference wave on the camera and created a modulated image, which could be processed through Hilbert transformation into the complex amplitude of the output speckle. The image was down-sampled, as exemplified in the magnified inset, and rearranged into a column vector of $T_{fw}$, with rows and columns indexed by the output and input channel positions, respectively. Only a subset of $T_{fw}$ is shown. (b) For round-trip measurements, the camera at the distal side was replaced by a mirror, and the returning light was directed by a non-polarizing beamsplitter to the same camera for holographic recording. The complex image of the round-trip output speckle was down-sampled at the 695 positions of the input foci grid (inset), resulting in a square matrix. A subset of $T_{2x}$ is shown, the vertical dashed line indicates the vector arranged as image of the facet, and the yellow inset shows sampling locations as white markers. The color maps encode complex values, and the scale bars in the insets are 50 μm.

## 2.2 Inspecting and controlling the spatial degrees of freedom

We quantified the number of guided modes within the MMF by performing singular value decomposition (SVD) on measured $T_{fw}$ and $T_{2X}$, counting the singular values (SVs) above a threshold defined as 5% of the largest SV. As shown in Fig. 3, there are ~500 populated modes in $T_{fw}$ and ~450 in $T_{2X}$, but mode-dependent transmission loss is apparent. While the numbers are consistent with a theoretical maximum of 550, when inspecting the left singular vectors associated with decaying SVs, we find that the loss of guided power increases as a mode carries higher radial frequencies. $T_{2x}$ exhibits more severe loss due to the double passage through the MMF, but loss may also be due to coupling and the detection of a single polarization state.

To show that the measured TMs are accurate, we controlled the amplitude and phase of the illumination wave front to physically create a sharp optical focus through the MMF on either the distal or the proximal end, using the measured $T_{fw}$ and $T_{2X}$, respectively. This was accomplished by numerically inverting the TMs and generating the required phase and amplitude pattern for an intended focusing position with the SLM using the Gerchberg-Saxton (GS) algorithm [25]. Because $T_{fw}$ is non-square, and both matrices are corrupted by noise and close to singular, we approximated matrix inversion with Tikhonov regularization, with the regularization parameter, $\gamma$, chosen as 10% of the greatest SV. This is justified based on the L-curve method [26]. The product of $T_{fw}$ with its regularized inverse is identical to the multiplication of a modified TM with its Hermitian transpose. The modification consists of rescaling each SV, $\sigma$, of the TM by $1/\sqrt{\sigma^2 + \gamma^2}$, and is shown in dashed curves (labeled as "regularized") in Fig. 3. Examples of the V polarization of the created foci are shown in the bottom row of Fig. 3, with ~4 µm average FWHM. We defined a focus contrast (FC) as the ratio of the peak intensity at the focal point over the average intensity across all output spatial channels to evaluate the focusing performance. This FC metric is similar to the enhancement factor defined by Vellekoop et al. [27,28], but it is bounded by the number of guided modes in the MMF even in a lossless condition and expresses accurately what fraction of the DOFs is effectively controlled. The average FCs for distal and proximal focusing were ~205.7 and ~148.3, respectively. Whereas the maximal FC, calculated when assuming the total power is concentrated in a single output spatial channel, would be 550 given the theoretical number of modes per polarization, the experimental FCs are limited by several factors such as the MMF loss, finite persistence time of the system, the measurement noise, the imperfect wave-front shaping, and the finite camera dynamic range. Despite the discrepancy between the experimental and theoretical values, the achieved FCs agree with the quantified number of modes, suggesting that we were reasonably exploiting the available DOF.

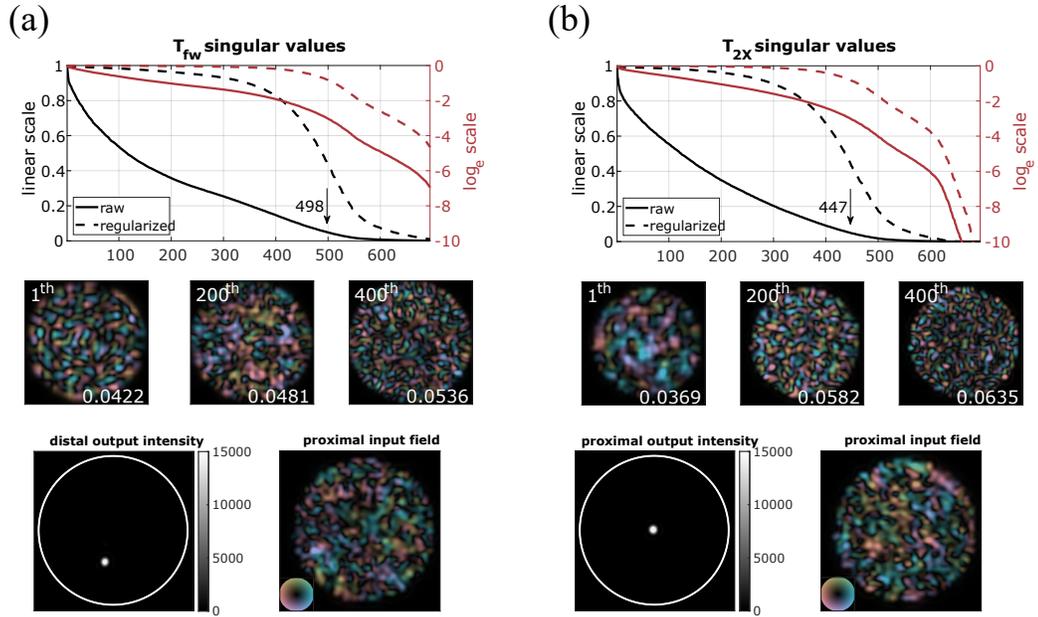

**Fig. 3.** Singular values (SVs) of the measured TMs and focusing through the MMF with regularized inversions in (a) single-pass forward $T_{fw}$ and (b) double-pass round-trip $T_{2X}$. The black arrows indicate the number of modes with an SV above 5% of the TM's largest SV. The solid and dashed curves correspond to raw and regularized SVs, while black and red lines show linear and log scales, respectively. Three examples of singular modes are visualized for each configuration by reshaping singular vectors to 2D images and numerically interpolating the images for better visual appearance. The averaged normalized radial frequency (0.5 cycles/radius) of the power spectral density of each mode is indicated in the lower-right corner. High-order modes are associated with higher radial frequency and are subject to increased loss. As shown in the bottom row, with the knowledge of $T_{fw}$ and $T_{2X}$, we can focus through the MMF on either distal or proximal facet at an intended position by tailoring the illumination wave front at the proximal end. The white circles outline the fiber facet, and the scale bar is 50 µm.

### 2.3 Transpose symmetry in round-trip transmission

Next, in a subsequent experiment, we set out to verify the anticipated transpose symmetry within the round-trip TM $T_{2X}$, as stated in Eq. 5. This property should be self-sustained and independent of $T_{fw}$. Physical misalignment between the defined input surface to the MMF and the image recording plane at the proximal end introduces a phase mismatch and relative shifts that need to be compensated to reveal the underlying transpose symmetry. This is similar to misalignment issues in common DOPC systems [29]. We parameterized the physical misalignment considering 8 variables and developed an optimization procedure that corrects the alignment imperfections, similar to Plöschner et al. [30]. To address the phase mismatch, we applied a two-dimensional (2D) phase term constituted by Zernike modes in the recording space of the output spatial channels. This corresponds to a diagonal phase-only matrix left-multiplied with $T_{2X}$. The Zernike orders correspond to 2D tilts, defocus, and 2D astigmatisms. To register the positional shifts, we applied another phase term with 2D tilts and defocus in the Fourier space of the output spatial channels of $T_{2X}$. This correction is equivalent to convolving the output spatial channels with a complex and offset point spread function. In the TM formalism, this is a further left-multiplication of $T_{2X}$ with a Toeplitz matrix. The Zernike coefficients were determined by minimizing the error $|T'^T_{2X} - T'_{2X}|^2$, where $T'_{2X}$ is the corrected $T_{2X}$, and $|\cdot|^2$ is the squared Frobenius matrix norm. Without correction, the initial error, normalized by $|T_{2X}|^2$, was 200%. Visualized in Fig. 4, the product of the uncorrected $T_{2X}$ with its regularized matrix inversion $T^{T-1}_{2X}$ is a chaotic matrix due to the disordered interference between populated modes caused by the physical misalignment. With correction, the normalized error was reduced to 23%. For comparison, we found a 15% residual error when computing the normalized squared Frobenius norm of the difference between two sequentially measured round-trip TMs of the identical MMF transmission. Moreover, the product of $T'_{2X}$ with $T'^{T-1}_{2X}$ became close to the identity matrix, with the integrated on-diagonal energy over the total matrix energy improving from 0.24% to 43.5%. As benchmark, the same metric applied to a perfectly symmetric TM, $(T'^T_{2X} + T'_{2X})/2$, resulted in 59.2% on-diagonal energy, limited by the regularized matrix inversion. These results show that the phase-corrected round-trip TM matches its transpose, thus demonstrating its transpose symmetry.

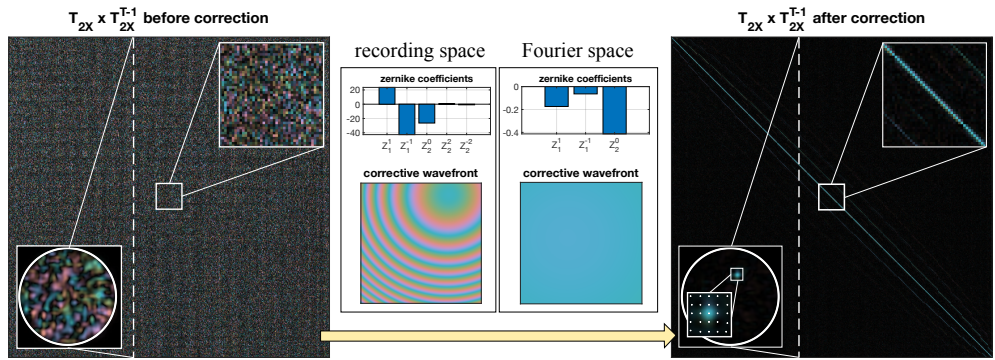

**Fig. 4.** Transpose symmetry of $T_{2X}$ before and after correction for misalignment at the proximal end. The phase mismatch and positional shifts are described as Zernike modes in the recording and Fourier space, respectively, and the amplitude of each mode is iteratively updated to minimize the difference between $T'_{2X}$ and $T'^T_{2X}$. Higher order Zernike modes are found to be negligible after additional trials. Using a Newtonian-based optimizer, the normalized error converges from 200% to 23% after the phase mismatch and positional shift corrections, each within tens of iterations. The horizontal and vertical tilts and defocus in the recording space are the dominant factors. A column in the products of $T_{2X}$ and $T^{T-1}_{2X}$ before and after phase correction is selected, converted back into 2D coordinates, and smoothed by interpolation to illustrate constructive interference at the corresponding proximal spatial channel when using corrected TMs. The offset diagonals on both sides of the main diagonal are due to oversampling during TM measurement, as visualized by indicating the proximal sampling positions.

## 2.4 Transpose relationship between the backward and forward transmission

With the corrected $T_{2X}$, we proceeded to verify the transpose relationship between the forward and backward TMs, as stated in Eq. 3. For experimental convenience, instead of directly comparing $T_{bw}$ and $T_{fw}$, we assumed that $T_{bw} = T^T_{fw}$ and worked with $T^T_{fw} \times T_{fw}$ and $T'_{2X}$, avoiding the complexity of directly measuring $T_{bw}$. Similar to correcting the round-trip measurements, we had to compensate the physical misalignment between the recording plane on the distal side for measuring $T_{fw}$ and the gold-coated mirror used in measuring $T_{2X}$. In a similar way to how we corrected $T_{2X}$, we applied phase terms to the recording and Fourier spaces of the output spatial channels of $T_{fw}$. In this case, we aimed to minimize the error $|T'_{2X} - T'^T_{fw} \times T'_{fw}|^2$, where $T'_{fw}$ is the corrected $T_{fw}$. Fig. 5 shows that the misalignment, characterized by the amplitude of the Zernike modes, was quite different from that encountered in $T_{2X}$. Without correction to $T_{fw}$, the initial error, normalized by $|T'_{2X}|^2$, was 101%, and the product of $T'_{2X}$ and $(T^T_{fw} \times T_{fw})^{-1}$ appeared far from a diagonal matrix, implying low resemblance between $T'_{2X}$ and $T^T_{fw} \times T_{fw}$. Clearly, the random background denotes that the physical misalignment caused undesired interference over all spatial channels. Crucially, the normalized error reduced to 27.7% after correction, which is again close to the experimental benchmark of 15%. Additionally, the resultant product closely resembled the identity matrix, with its integrated on-diagonal energy over the total matrix energy improving from 0.27% to 36.6%. The product of $T'_{2X}$ with its regularized inverse reached 56.8% on-diagonal energy. Therefore, we conclude that $T'_{2X}$ and $T'^T_{fw} \times T'_{fw}$ are identical to each other, as stated in

Eq. 4., which implies that the backward transmission $T_{bw}$ is the same as $T^T_{fw}$, as described in Eq. 3. This provides a convincing proof of general optical reciprocity and the ensuing transpose symmetry for transmission through a MMF, which serves as a convenient model for general complex media.

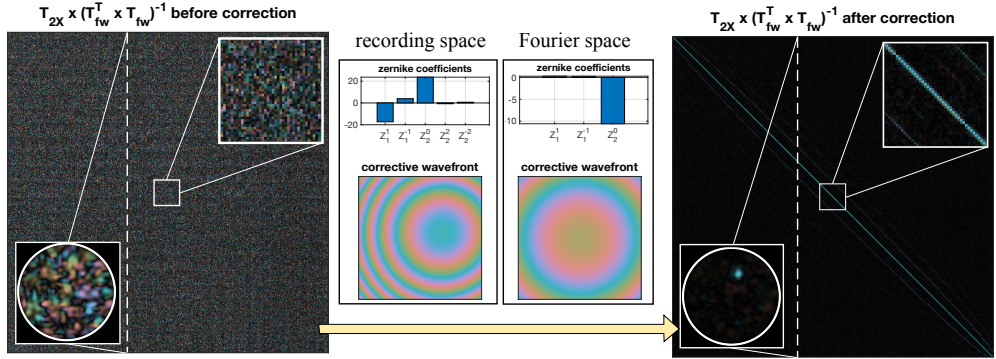

**Fig. 5.** Visualization of optical reciprocity within the MMF after correcting $T_{fw}$ for misalignment. During the optimization, the amplitude of each Zernike mode is iteratively updated to minimize the difference between experimental and synthesized round-trip transmission. The Newtonian-based optimizer was again used to find the optimal correction. The on-diagonal energy ratio improved from 0.27% to 36.6% after the phase mismatch and positional shift corrections, each within tens of iterations.

### 2.5 Optical phase conjugation based on time reversibility in reciprocal medium

DOPC based on time reversibility instructs that light propagation can be reversed along its pathway by conjugating its field. In the MMF, a given propagation pathway from distal input to proximal output can be retraced by a proximal input conjugate to the proximal output, resulting in a distal output in the same spatial channels as the original distal input. This process is represented by substituting $\vec{s}$ in Eq. 1 with the complex conjugate of Eq. 2. Applying Eq. 3, the distal output field (we use $\vec{v}$ rather than $\vec{t}$ to avoid symbol confusion) becomes

$$\vec{v} = T_{fw} \times T^{\dagger}_{fw} \times \vec{t}^{*}, \tag{6}$$

where the superscript * indicates complex conjugation, and the superscript † indicates Hermitian transpose. For a unitary linear system without loss, the Hermitian transpose of $T_{fw}$ equals its inverse $T^{\dagger}_{fw} = T^{-1}_{fw}$, and Eq. 6 reduces to $\vec{v} = \vec{t}^{*}$. As a result, we can reproduce at the distal end the conjugated wave-front of any initial $\vec{t}$ through the MMF. The experimental realization of Eq. 6 is demonstrated in Fig. 6(a), where we replicated a diffraction-limited focal spot through the MMF with DOPC. The laser, optics, and camera were identical to those in Fig. 2. A 2.5 μm focal spot in the V polarization state was coupled into the MMF through a distal spatial channel. The resulting proximal field was recorded by off-axis holography. We then configured the SLM to send a conjugated copy of the recorded wave-front in the same polarization state back into the MMF from the proximal side. The phase-conjugate light field retraced the forward light propagation and reconstructs the distal focal spot in the V polarization state with a diffraction-limited FWHM of ~4 μm and an FC of 91.4 at the original focusing position. Imperfect FC (below 550) of the MMF-generated focus can be attributed to losses in the MMF and the measurement system, which violate the power conservation in time reversibility and lead to an only approximate time reversal symmetry in the MMF light transport process. Note that higher FCs of generated foci were obtained with regularized matrix inversion.

To investigate how loss influences the DOPC-based focusing performance, we simulated different loss conditions by replacing the SVs of experimentally measured TMs with exponential functions that have varying decay constants and are multiplied by a step function with a cutoff at 550. A decay constant of zero corresponds to a sharp step function. To include the effect of measurement noise and experimental limitations, we employed two replicate measurements $T_{fw}$ and $\widetilde{T}_{fw}$ of an identical MMF transmission, where the singular vectors of $T_{fw}$ and $\widetilde{T}_{fw}$ are noisy. The apparent loss was quantified as percentage decrease in the square of the Frobenius norm of the TM. For each loss condition, we computed the Hadamard product $(T_{fw} \times \widetilde{T}^{\dagger}_{fw})^{\circ 2}$, which is the element-wise square of $T_{fw} \times \widetilde{T}^{\dagger}_{fw}$, to simulate MMF focusing and obtain the averaged FC by calculating the averaged ratio of each on-diagonal to the mean value of the corresponding column in the Hadamard product. As shown in the solid curve in Fig. 6(b), the FC declines with increasing loss, resulting in FCs of 491.5, 162.4, and 19.8 at losses of 0, 85, and 98%, corresponding to the no loss, equivalent to experimental loss, and substantial loss conditions, respectively. 2D images of example simulated DOPC-based MMF focusing at the conditions were obtained by reshaping a select column of $(T_{fw} \times \widetilde{T}^{\dagger}{}_{fw})^{o2}$, revealing a prominent background with high loss. Using, instead, the Tikhonov-regularized matrix inversion approach, as described in Section 2.2, to compute $(T_{fw} \times \widetilde{T}^{-1}{}_{fw})^{o2}$ improved the MMF focusing and FCs (dashed curve in Fig. 6(b)) of 491.5, 356.1, and 49.3, in the three highlighted conditions, respectively. Fig. 6(b) also shows the experimentally achieved FCs of 205.7 from Section 2.2 and 91.4 from DOPC here and their apparent loss for comparison. Due to the imperfect wave-front shaping and limited camera dynamic range, the experimental FCs are inferior.

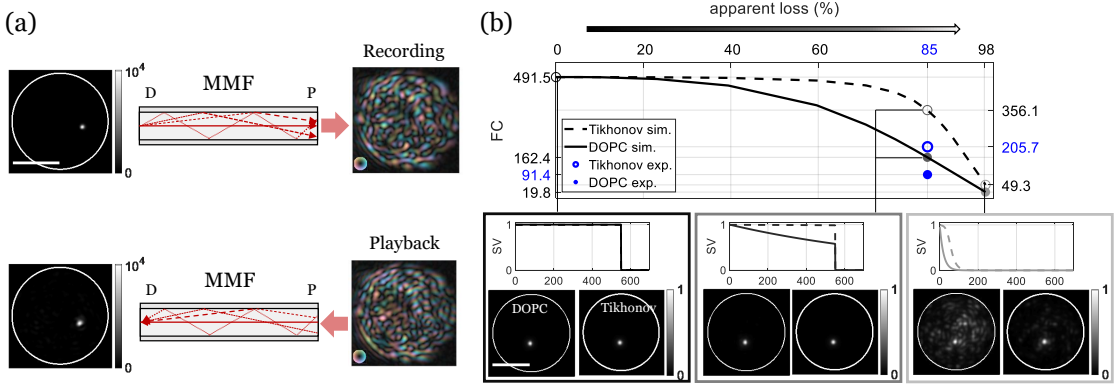

**Fig. 6.** (a) Focusing through a MMF based on time-reversibility with experimental DOPC. D: distal end. P: proximal end. In the recording phase, a laser light is focused on the MMF distal facet, and the output speckle field is interferometrically recorded. In the playback phase, the conjugated copy of the recorded wave-front is projected on the proximal facet. This reproduces the distal focal spot, which is clearly visible at the original focusing position and has a FWHM slightly larger due to the limited NA of the MMF. (b) Simulated DOPC- and Tikhonov-inversion-based MMF focusing using $T_{fw}$ with varying apparent loss. Specific focus examples as well as their corresponding SV distributions (solid curves) are displayed. While the focus remains clearly visible as the background signal increases with loss, Tikhonov regularized inversion compensates to some extent for the mode-dependent loss (dashed curves) and improves the FCs, depending on the loss condition. The white circles outline the fiber facet, and the scale bars are 50 μm.

### III. Discussion

Optical reciprocity is a universal principle within linear, non-magnetic, and static media, even in the presence of loss. It has been previously shown in various formalisms and contexts [1,2,31,32] and extends to complex media such as a MMF. We investigated symmetry constraints that reciprocity imposes on bi-directional light transport through a MMF and measured the forward and double-pass TMs, $T_{fw}$ and $T_{2X}$, to demonstrate that $T^{T}{}_{2X} = T_{2X}$ and $T_{2X} = T^{T}{}_{fw} \times T_{fw}$. The round-trip transmission reveals a transpose symmetry, and the backward transmission presents a transpose relationship to the forward transmission. We also showed that focusing through a MMF with DOPC, which relies on time reversibility by assuming a loss-free transmission, may have limitation when in practice the transmission suffers from non-negligible loss. This means that while time reversibility is a consequence of optical reciprocity, not all reciprocal systems allow time reversal.

When the MMF is truly loss-free, and we have a fully sampled $T_{fw}$ with uniform SVs, the number of degrees of control is the same as the number available DOF. In this case, $T_{fw} \times T^{\dagger}{}_{fw}$ is an low-pass filtered identity matrix, suggesting that we can focus through the MMF based on DOPC and achieve an FC close to the number of guided modes. Noise is the only limiting factor, and DOPC- and Tikhonov-inversion based MMF focusing have the same performance. In our experimental condition, the MMF may leak some of the modes, and only measuring a single polarization state intrinsically eliminates all power in the orthogonal polarization states. Nevertheless, using DOPC we could still generate a focal spot with an FC of 91.4 through a lossy MMF, treating the transmission as an approximately unitary system, for which $T^{\dagger}{}_{fw} \sim T^{-1}{}_{fw}$. On the other hand, experimental focusing through the MMF with regularized TM inversion achieved a better FC of more than 200. This is because the Tikhonov regularization numerically compensates the mode-dependent loss and creates a balanced constructive interference, providing a better focusing performance. However, if the MMF transmission is dissipative, $T_{fw} \times T^{\dagger}{}_{fw}$ is far from an identity matrix, and the generated focus with DOPC barely stands out from the speckle background. In this regime, using Tikhonov regularization may only have modest benefit, since it only compensates for SVs experiencing modest loss. The DOF corresponding to SVs smaller than the regularization parameter remain uncontrolled and do not contribute to the constructive interference at intended focus locations.

The implementation of a flexible MMF endoscope remains technically challenging despite recently proposed strategies [30,33–36], and the lack of flexibility is the enduring bottleneck for MMF imaging applications. Because the TMs of MMFs are notoriously sensitive to physical fiber deformation [30], a flexible MMF endoscope would demand repeated on-site calibration without open distal access in practical endoscopic settings. Although calibrating a MMF with only proximal access is a desirable strategy, robust experimental MMF proximal calibration methods remain to be demonstrated. Understanding the reciprocal nature of light propagation through a MMF and the underlying symmetry constraints may help tackle this challenge. In the context of proximal MMF calibration, where measurement of $T_{2X}$ may be available, the demonstrated symmetry constraint precludes straightforward recovery of $T_{fw}$ or $T_{bw}$, which is needed for imaging through the MMF [15,30,37,38]. To appreciate this limitation, we can factor $T_{fw}$ into its symmetric and anti-symmetric part based on the second polar decomposition [39],

$$T_{fw} = A \times L, \qquad (7)$$

where A is orthogonal ($A^T = A^{-1}$), and L is transpose symmetric ($L^T = L$). In this case Eq. 4 becomes

$$T_{2X} = L^2. \qquad (8)$$

The orthogonal parts cancel each other upon forward and backward transmission, preserving only the symmetric part in the round-trip transmission measurement. Eq. 8 states a fundamental restriction: while the symmetric part of $T_{fw}$ can be uniquely retrieved by taking the matrix square-root of the proximally measured $T_{2X}$ [40], if it has no negative real eigenvalues, the orthogonal part, A, vanishes due to the intrinsic propagation property imposed by the optical reciprocity. Put differently, although a square, complex-valued matrix of dimension $M$ has $2M^2$ unknown coefficients, the transpose symmetry reduces this number to $M^2 + M$, masking the additional $M^2 - M$ of the orthogonal component. This leads to symmetric degeneracy of $T_{fw}$ even though $T_{2X}$ is known. This explains why $T_{fw}$ cannot be directly retrieved from $T_{2X}$, which complicates strategies for MMF proximal calibration methods.

Previously, Takagi matrix factorization has been proposed to help in recovering $T_{fw}$ from proximal measurements [33]. A carefully engineered static reflector installed at the distal end of a MMF can provide distinctive reflectivity on individual distal spatial channels, which augments Eq. 4 to

$$T_{2X} = T^T_{fw} \times R \times T_{fw}, \qquad (9)$$

where R is a real-valued diagonal matrix with sortable on-diagonal elements. By performing Takagi factorization on the accessible $T_{2X}$ and leveraging the transpose symmetry, we have

$$T_{2X} = U^T \times \Sigma \times U, \qquad (10)$$

where U is a unitary matrix, and $\Sigma$ is a real-valued diagonal matrix. If the MMF is loss-free, this would suggest $R = \Sigma$ and $T_{fw} = D \times U$, where D is an unknown diagonal matrix with entries that are ± 1 and might be estimated with prior knowledge. Unfortunately, as shown in Fig. 3, $T_{fw}$ is generally lossy, which compromises this strategy. Based on our findings and arguments, breaking the intrinsic transpose symmetry in the round-trip transmission, installing a calibration element capable of several realizations at the MMF distal end, or introducing new constraints by measuring multispectral round-trip transmission to resolve the degeneracy issue might be the most viable solution towards a flexible MMF endoscope [6,33,41]. Since reciprocity is ubiquitous, the symmetry principle may also inform non-invasive imaging, coherent wave-control through highly scattering tissues, and electromagnetic communications.

In conclusion, optical reciprocity imposes a symmetry on the bi-directional propagation through a general complex medium regardless of the path complexity or loss. We experimentally demonstrated this symmetry in a looped 1-m-long step-index MMF by measuring the forward and round-trip transmissions. The symmetry prohibits direct retrieval of the forward TM from a round-trip measurement. Thus MMF endoscopy in a practical setting is fundamentally complicated due to the need to calibrate the MMF without distal access. The insights of light transport within a MMF obtained here may stimulate improved strategies for flexible MMF endoscopy and facilitate efficient sensing and imaging techniques through complex or disordered media.

## IV. Acknowledgements

Research in this publication was supported by the National Institute of Biomedical Imaging and Bioengineering of the National Institutes of Health, award P41 EB015903. VJP was supported by the OSA Deutsch fellowship.

V. Data availability

The data that support the findings of this study are available from the corresponding author upon reasonable request.

**References:**

1. H. A. Haus, *Waves and Fields in Optoelectronics* (Prentice-Hall, 1984).
2. R. J. Potton, "Reciprocity in optics," Reports Prog. Phys. **67**(5), 717–754 (2004).
3. L. Onsager, "Reciprocal Relations in Irreversible Processes. I.," Phys. Rev. **37**(4), 405–426 (1931).
4. H. B. G. Casimir, "Reciprocity Theorems and Irreversible Processes," Proc. IEEE **51**(11), 1570–1573 (1963).
5. M. G. Silveirinha, "Hidden time-reversal symmetry in dissipative reciprocal systems," Opt. Express **27**(10), 14328 (2019).
6. Y. Bromberg, B. Redding, S. M. Popoff, and H. Cao, "Control of coherent backscattering by breaking optical reciprocity," Phys. Rev. A **93**(2), 1–16 (2016).
7. D. S. Wiersma, "Breaking reciprocity," Nat. Photonics **6**(8), 506–507 (2012).
8. M. Cui and C. Yang, "Implementation of a digital optical phase conjugation system and its application to study the robustness of turbidity suppression by phase conjugation," Opt. Express **18**(4), 3444–3455 (2010).
9. I. N. Papadopoulos, S. Farahi, C. Moser, and D. Psaltis, "Focusing and scanning light through a multimode optical fiber using digital phase conjugation," Opt. Express **20**(10), 10583–10590 (2012).
10. Z. Yaqoob, D. Psaltis, M. S. Feld, and C. Yang, "Optical phase conjugation for turbidity suppression in biological samples," Nat. Photonics **2**(2), 110–115 (2008).
11. I. M. Vellekoop, M. Cui, and C. Yang, "Digital optical phase conjugation of fluorescence in turbid tissue," Appl. Phys. Lett. **101**(8), 81108 (2012).
12. A. P. Mosk, A. Lagendijk, G. Lerosey, and M. Fink, "Controlling waves in space and time for imaging and focusing in complex media," Nat. Photonics **6**(5), 283–292 (2012).
13. D. Psaltis and C. Moser, "Imaging with Multimode Fibers," Opt. Photonics News (January), 14–16 (2010).
14. D. J. Richardson, J. M. Fini, and L. E. Nelson, "Space-division multiplexing in optical fibres," Nat. Photonics **7**(5), 354–362 (2013).
15. Y. Choi, C. Yoon, M. Kim, T. D. Yang, C. Fang-Yen, R. R. Dasari, K. J. Lee, and W. Choi, "Scanner-free and wide-field endoscopic imaging by using a single multimode optical fiber," Phys. Rev. Lett. **109**(20), 1–5 (2012).
16. T. Čižmár and K. Dholakia, "Exploiting multimode waveguides for pure fibre-based imaging," Nat. Commun. **3**, (2012).
17. P. Caramazza, O. Moran, R. Murray-Smith, and D. Faccio, "Transmission of natural scene images through a multimode fibre," Nat. Commun. **10**(1), 2029 (2019).
18. S. Li, S. A. R. Horsley, T. Tyc, T. Cizmar, and D. B. Phillips, "Guide-star assisted imaging through multimode optical fibres," (2020).
19. S. M. Popoff, G. Lerosey, R. Carminati, M. Fink, A. C. Boccara, and S. Gigan, "Measuring the Transmission Matrix in Optics: An Approach to the Study and Control of Light Propagation in Disordered Media," Phys. Rev. Lett. **104**(10), 100601 (2010).
20. S. Popoff, G. Lerosey, M. Fink, A. C. Boccara, and S. Gigan, "Image transmission through an opaque material," Nat. Commun. **1**(1), 81 (2010).
21. S. M. Popoff, G. Lerosey, R. Carminati, M. Fink, A. C. Boccara, and S. Gigan, "Measuring the transmission matrix in optics: An approach to the study and control of light propagation in disordered media," Phys. Rev. Lett. **104**(10), 1–4 (2010).
22. D. Jalas, A. Petrov, M. Eich, W. Freude, S. Fan, Z. Yu, R. Baets, M. Popović, A. Melloni, J. D. Joannopoulos, M. Vanwolleghem, C. R. Doerr, and H. Renner, "What is — and what is not — an optical isolator," Nat. Photonics **7**(8), 579–582 (2013).
23. A. Yariv and P. Yeh, *Photonics : Optical Electronics in Modern Communications* (Oxford University Press, 2007).
24. W. Xiong, C. W. Hsu, Y. Bromberg, J. E. Antonio-Lopez, R. Amezcua Correa, and H. Cao, "Complete polarization control in multimode fibers with polarization and mode coupling," Light Sci. Appl. **7**(1), 54 (2018).
25. G. Yang, B. Dong, B. Gu, J. Zhuang, and O. K. Ersoy, "Gerchberg–Saxton and Yang–Gu algorithms for phase retrieval in a nonunitary transform system: a comparison," Appl. Opt. **33**(2), 209–218 (1994).
26. P. C. Hansen, *The L-Curve and Its Use in the Numerical Treatment of Inverse Problems* (WIT Press, 2000).
27. I. M. Vellekoop and A. P. Mosk, "Universal Optimal Transmission of Light Through Disordered Materials," Phys. Rev. Lett. **101**(12), 120601 (2008).
28. I. M. Vellekoop, "Feedback-based wavefront shaping," Opt. Express **23**(9), 12189 (2015).
29. M. Jang, H. Ruan, H. Zhou, B. Judkewitz, and C. Yang, "Method for auto-alignment of digital optical phase conjugation systems based on digital propagation," Opt. Express **22**(12), 14054 (2014).
30. M. Plöschner, T. Tyc, and T. Čižmár, "Seeing through chaos in multimode fibres," Nat. Photonics **9**(8), 529–535 (2015).
31. Z. Sekera, "Scattering Matrices and Reciprocity Relationships for Various Representations of the State of Polarization," J. Opt. Soc. Am. **56**(12), 1732–1740 (1966).
32. P. P. Khial, A. D. White, and A. Hajimiri, "Nanophotonic optical gyroscope with reciprocal sensitivity enhancement," Nat. Photonics **12**(11), 671–675 (2018).
33. R. Y. Gu, R. N. Mahalati, and J. M. Kahn, "Design of flexible multi-mode fiber endoscope," Opt. Express **23**(21), 26905 (2015).
34. R. Y. Gu, E. Chou, C. Rewcastle, O. Levi, and J. M. Kahn, "Improved spot formation for flexible multi-mode fiber endoscope using partial reflector," **200**(2015), 23845–23858 (2018).
35. A. M. Caravaca Aguirre, E. Niv, D. B. Conkey, and R. Piestun, "Real time focusing through a perturbed multimode fiber," Imaging Appl. Opt. CTh2B.4 (2013).
36. S. Farahi, D. Ziegler, I. N. Papadopoulos, D. Psaltis, and C. Moser, "Dynamic bending compensation while focusing through a multimode fiber," Opt. Express **21**(19), 22504 (2013).
37. D. Loterie, S. Farahi, I. Papadopoulos, A. Goy, D. Psaltis, and C. Moser, "Digital confocal microscopy through a multimode fiber," Opt. Express **23**(18), 23845 (2015).
38. N. Borhani, E. Kakkava, C. Moser, and D. Psaltis, "Learning to see through multimode fibers," Optica **5**(8), 960–966 (2018).
39. R. Bhatia, "The bipolar decomposition," Linear Algebra Appl. **439**(10), 3031–3037 (2013).

40. C. R. Johnson, K. Okubo, and R. Reams, "Uniqueness of matrix square roots and an application," **323**, 51–60 (2001).
41. G. S. D. Gordon, M. Gataric, A. G. C. P. Ramos, R. Mouthaan, C. Williams, J. Yoon, T. D. Wilkinson, and S. E. Bohndiek, "Characterizing Optical Fiber Transmission Matrices Using Metasurface Reflector Stacks for Lensless Imaging without Distal Access," Phys. Rev. X **9**(4), 41050 (2019).